\documentclass{mem}
\usepackage{natbib}\usepackage{txfonts}\usepackage{balance}
\usepackage{flushend}
\usepackage{xcolor}
\usepackage{graphicx}
\usepackage[a4paper,breaklinks,pdftex]{hyperref}
\idline{75}{282}
\begin{document}
\def\teff{$T\rm_{eff }$}
\def\kms{$\mathrm {km s}^{-1}$}

\title{
Current and Future constraints on Very-Light Axion-Like Particles from X-ray observations of cluster-hosted Active Galaxies}

   \subtitle{}

\author{
J.M. \,Sisk-Reynes,\inst{1} 
C.S. \,Reynolds\inst{1}
\and J.H. \,Matthews\inst{2}
}

\institute{Institute of Astronomy, University of Cambridge, Madingley Rd, Cambridge, CB3 0HA\and 
Department of Physics, Astrophysics, University of Oxford, Denys Wilkinson Building, Keble Road, Oxford OX1 3RH
\\ 
\email{jms332@ast.cam.ac.uk}
}

\authorrunning{Sisk-Reynes}

\titlerunning{Astrophysical X-ray Very-Light ALP constraints}

\date{Received: 4 December 2022; Accepted: Day Month Year}

\abstract{We discuss our recent constraints on the coupling of Very-Light Axion-Like Particles (of masses $<$$ 10^{-12} \ \mathrm{eV}$) to electromagnetism from \textit{Chandra} observations of the cluster-hosted Active Galactic Nuclei (AGN)~H1821+643 and NGC1275.~In both cases, the inferred high-quality AGN spectra excluded all photon-ALP couplings $g_\mathrm{a\gamma} > (6.3 - 8.0) \times 10^{-13} \ {\mathrm{GeV}}^{-1}$ at the $99.7\%$ level, respectively, based on the non-detection of spectral distortions attributed to photon-ALP inter-conversion along the cluster line-of-sight.~Finally, we present the prospects of tightening current bounds on such ALPs by up to a factor of 10 with next-generation X-ray observatories such as \textit{Athena}, \textit{AXIS} and \textit{LEM} given their improved spectral and spatial resolution and collecting area compared to current missions.
}
\maketitle{}

\section{Overview} Axion-Like Particles (ALPs) are model-independent generalisations of the Quantum Chromodynamics axion, generic predictions of string theories \citep{Conlon_2006_iib,Svrcek_witten_2006_iib,greenSchwarzWitten_2012,Cicoli_2012_typeiib} and Dark Matter candidates \citep{PRESKILL_Wise_Wilczek_CosmoInvisibleAxion_83,Abbott_Sikivie_CosmologicalBoundOnAxion_83,Dine_Fischler_83_HarmlessAxion}.\\ \\ \indent Astronomy provides rich and magnetised environments that probe the interaction of ALPs with electromagnetism as characterised by the coupling strength $g_\mathrm{a\gamma}$ for a given ALP mass $m_\mathrm{a}$.~For a photon beam passing through a magnetised plasma $\textbf{B}$, the photon-ALP interaction is described by the Lagrangian density: \begin{equation}
    \label{equation:eq_lagrangian}
    \mathcal{L}_\mathrm{a\gamma} = g_\mathrm{a\gamma} ~ a ~ \textbf{E} \cdot \textbf{B}
    \centering 
\end{equation}where $\textbf{E}$ and $a$ are the electric and ALP fields.\begin{figure*}[]
{\includegraphics[width=.9\textwidth,height=0.7\textwidth]{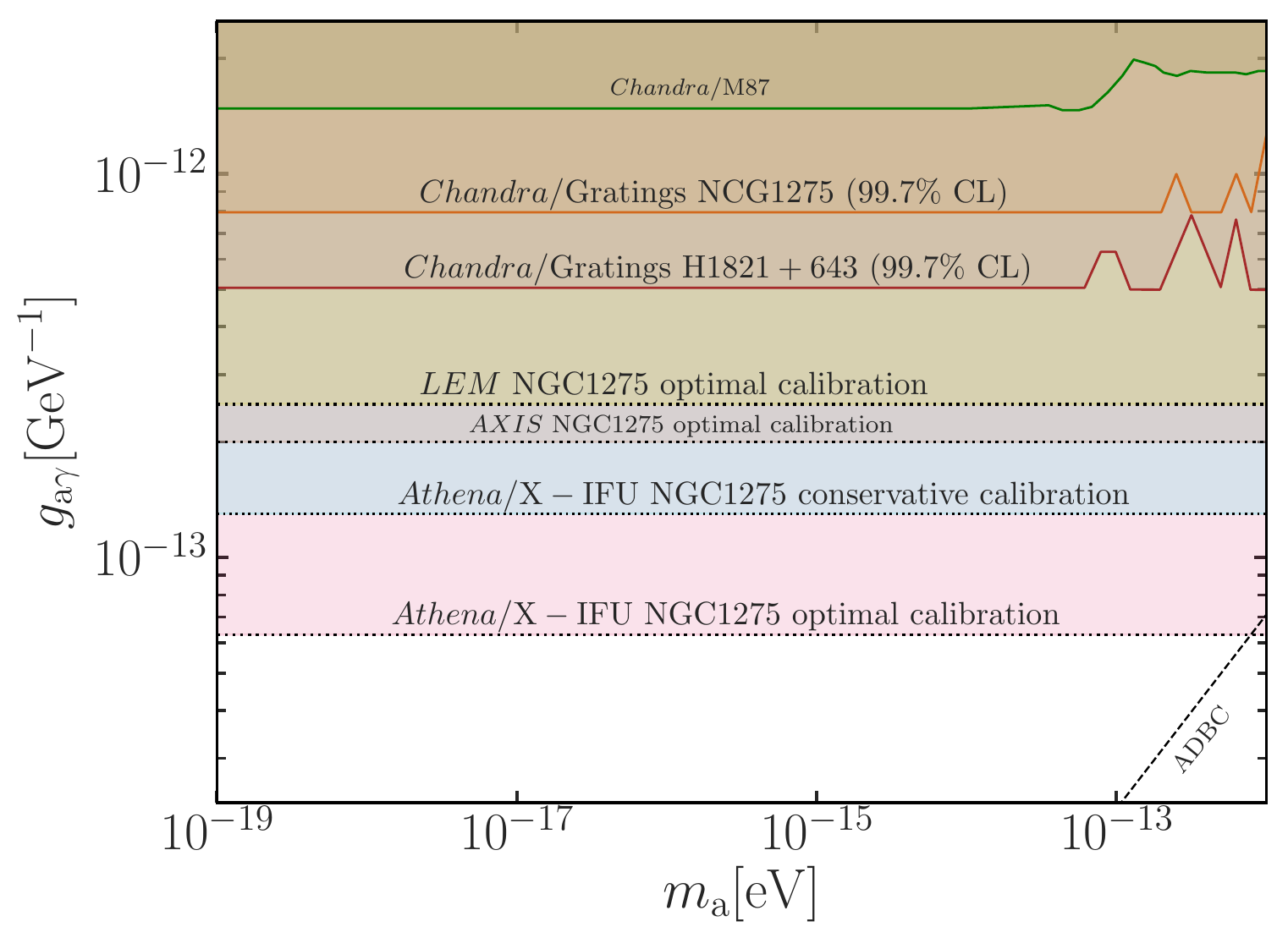}}
\caption{\footnotesize
Constraints on Very-Light ALPs from \textit{Chandra} observations of cluster-hosted AGN.~We include projections from the future: \textit{Athena}, \textit{AXIS} and \textit{LEM} X-ray observatories; and the ground-based birefringent cavity for the study of ALP Dark Matter \textit{ADBC}.~All results are quoted at the $95\%$ level unless specified \citep[adapted from][]{siskreynes_22_nextGen}.}
\label{figure:results_with_lem}
\end{figure*} 
Equation \ref{equation:eq_lagrangian} illustrates that, for a suitable location of parameter space, photon-ALP mixing will take place as photons from a point source traverse a magnetised plasma.~For a fixed set of ALP parameters $(m_\mathrm{a}, g_\mathrm{a\gamma}$), this effect will induce energy-dependent distortions, or absorption-like features, in the source spectrum.~The presence or absence of such distortions can be used to constrain $g_\mathrm{a\gamma}$.

Galaxy clusters are the most massive bound objects in the Universe and are permeated by the Intracluster Medium (ICM).~The ICM is magnetised and turbulent over scales of $100 \ \mathrm{pc} - 10 \ \mathrm{kpc}$ and can attain attain field strengths up to 10\,$\mu$G~in the cores of the richest (dynamically relaxed) clusters \citep{govoni+ferreti04}.~Some of such cool-core clusters host an Active Galactic Nucleus (AGN) at their Brightest Cluster Galaxy.~High-quality spectroscopic observations of cluster-hosted AGN taken with X-ray telescopes can provide a clear view of the intrinsic AGN emission free from ICM contamination~(e.g.~emission lines),~which can~be used to constrain ALPs.

\section{Current constraints on Very-Light ALPs from cluster-hosted AGNs}

The tightest constraints to date on Very-Light ALPs were inferred with a set of deep \textit{Chandra} Transmission Grating observations ($570$-ks of total exposure) of the very-luminous radio-quiet cluster-hosted quasar H1821+643 \citep{siskreynes_22_h1821}.~These observations,~taken in 2001, disperse the AGN emission across the grating array and enable a high-quality extraction of the AGN spectrum free from photon pile-up and cluster contamination.~{\citet{siskreynes_22_h1821} first described the intrinsic AGN emission with an astrophysical model typical of type-1 AGN.~This ``non-ALP'' model, whose free parameters were found by minimising the fit statistic, yielded energy-dependent residuals only below the 2.5\%~level.~To assess whether these remaining spectral distortions could be described by AGN photons undergoing inter-conversion into ALPs along the cluster line-of-sight, the authors first generated a set of photon-ALP mixing curves across a wide range of ALP parameters, $(m_\mathrm{a}, g_\mathrm{a\gamma})$, for 500 possible cell-based realisations of the turbulent cluster magnetic field.~Each of these curves was then fitted to the data as a multiplicative term modifying the astrophysical model.~After marginalising over all field realisations for a given set of ALP parameters $(m_\mathrm{a}, g_\mathrm{a\gamma})$, the authors computed posteriors on ALPs by comparing the fit statistics of the ``non-ALP'' and ALP-containing models.~At the $99.7\%$ confidence level (99.7\% CL), this assessment excluded all photon-ALP coupling values $g_\mathrm{a\gamma} > 6.3\times 10^{-13} \ {\mathrm{GeV}}^{-1}$ for $m_\mathrm{a}$$<10^{-12} \ \mathrm{eV}$.

Nevertheless, the ALP constraints inferred from H1821+643 are limited by the assumption of a constant thermal-to-magnetic pressure ratio ($\beta_\mathrm{pl}$)~up to the cluster virial radius.~However, one would expect bounds on $g_\mathrm{a\gamma}$ to relax by only up to $0.3 \ \mathrm{dex}$ with a radially-dependent $\beta_\mathrm{pl}$ \citep{matthews22_bfields}.~Moreover, a more realistic and sophisticated field model could yield tighter constraints on $g_\mathrm{a\gamma}$ \citep{carenza_fields}.~The previous best constraints on light ALPs had been found by \cite{schallmoser21}
and \citet{reynolds20_ngc1275},~where the former employed machine learning to improve on earlier bounds \citep{conlon17_manysources}.~A high-quality \textit{Chandra} Grating view of the central engine of the Perseus cluster, NGC1275, combined with assumptions about the magnetic field similar to those we made for H1821+643, excluded all photon-ALP coupling values $g_\mathrm{a\gamma} > 8.0\times 10^{-13} \ {\mathrm{GeV}}^{-1}$ for $m_\mathrm{a} < 10^{-12} \ \mathrm{eV}$ \citep{reynolds20_ngc1275}.

\section{The importance of collecting area and spectral and spatial resolution}
\citet{siskreynes_22_nextGen} showed that the next-generation X-ray observatories \textit{Athena} and \textit{AXIS}, with their unprecendented collecting areas and superior spectral and angular resolution, respectively, will further exclude~$g_\mathrm{a\gamma} > (1 - 4)\times 10^{-13} \ {\mathrm{GeV}}^{-1}$  \citep[see also][]{conlon_athena}.~These results,~shown in Fig.~\ref{figure:results_with_lem},~were inferred by simulating observations of NGC1275 with moderate exposure (200-ks) to facilitate comparison with the \textit{Athena}/X-IFU projection found by~\citet{conlon_athena}.~Importantly, in \citet{siskreynes_22_nextGen},~the updated \textit{Athena} bounds~(Fig.~\ref{figure:results_with_lem})~were inferred under a machine learning assessment of detector mis-calibration,~given its potential to mimic spectral distortions that could be attributed to photon-ALP~inter-conversion.~Additionally, Fig.~\ref{figure:results_with_lem} shows the projected bound on Very-Light ALPs we infer from a deep (1-Ms)~observation of NGC1275 taken with the next-generation~\textit{Line Emission Mapper} observatory \citep[\textit{LEM},][for a target Half Power Diameter of $10\arcsec$]{lem}.~Clearly, with a high-quality view of the intrinsic AGN emission with an unprecendented spectral resolution (0.9 eV within 0.5-2 keV), as well as with its prominent collecting area, \textit{LEM}~will exclude all photon-ALP couplings $g_\mathrm{a\gamma} > 2.5 \times 10^{-13} \ {\mathrm{GeV}}^{-1}$ (95\% CL) for Very-Light ALPs.~Therefore, the \textit{Athena}, \textit{AXIS} and \textit{LEM} projections illustrated in Fig.~\ref{figure:results_with_lem} highlight the exciting prospects of constraining Very-Light ALPs with next-generation X-ray observatories with fundamentally different designs.
\section{Conclusions}
Bright AGN hosted by rich cool-core galaxy clusters are excellent probes of Very-Light ALPs.~Current and next-generation observatories, including the \textit{Chandra}, \textit{Athena}, \textit{AXIS} and \textit{LEM} missions have excluded and will exclude photon-ALP couplings down to $g_\mathrm{a\gamma}\sim 10^{-13} - 10^{-12} \ {\mathrm{GeV}}^{-1}$ for ALP masses $<$$10^{-12} \ \mathrm{eV}$ with high-quality observations of H1821+643 and NGC1275.~In future, probing ALPs in this regime may be the \textit{only} plausible observational test of string theories \citep{halverson19}. 

\begin{acknowledgements} We thank~Anna~Ogorzalek~for providing the \textit{LEM} responses, as well as the reviewer for insightful comments.\end{acknowledgements}


\begin{thebibliography}{}


\bibitem[{Conlon~(2006)}]{Conlon_2006_iib}
Conlon,~J.,~2006,~JHEP,~5,~78


\bibitem[{Svrcek,P.~et~al.~(2006)}]{Svrcek_witten_2006_iib}
Svrcek,~P.\ and~Witten,~E.,~2006,~JHEP,~6,~51

\bibitem[{Green~et~al.~(2012)}]{greenSchwarzWitten_2012}
Green,~M.~et~al.,~2012,~Cambridge~Monogr.~Math.\\ Phys,~ISBN:~978-0-521-35752-4

\bibitem[{Cicoli~et~al~(2012)}]{Cicoli_2012_typeiib}
Cicoli,~M.~et~al.,~2012,~JHEP,~10,~146

\bibitem[{Preskill~et~al~(1983)}]{PRESKILL_Wise_Wilczek_CosmoInvisibleAxion_83}
Preskill,~J.~et~al.,~1983,~PR.~B.,~120,~1

\bibitem[{Abbott~et~al~(1983)}]{Abbott_Sikivie_CosmologicalBoundOnAxion_83}
Abbott,~L.\ and~Sikivie,~P.,~1983,~PR.~B.,~120

\bibitem[{Dine~et~al.~(1983)}]{Dine_Fischler_83_HarmlessAxion}
Dine,~M.\ and Fischler,~W.,~1983,~PR.~B.,~120


\bibitem[{Govoni~et~al.~(2004)}]{govoni+ferreti04}
Govoni,~F.\ and~Ferreti,~L.,~2004,~Int.~J.~Mod.\\ Phys,~D13

\bibitem[{Sisk-Reynes~et~al.~(2022b)}]{siskreynes_22_nextGen} 
Sisk-Reynes,~J.~et~al.,~2022,~arXiv,~2211.05136

\bibitem[{Carenza~et~al.~(2022)}]{carenza_fields} 
Carenza,~P.~et al.,~2022,~arXiv,~2208.04333

\bibitem[{Sisk-Reynes~et~al.~(2022a)}]{siskreynes_22_h1821} Sisk-Reynes,~J.~et al.,~2022,~\mnras,
510,1
\bibitem[{Reynolds~et~al.~(2020)}]{reynolds20_ngc1275} Reynolds,~C.~et~al.,~2020,~\apj,~890,~1

\bibitem[{Conlon~et~al.~(2018)}]{conlon_athena} Conlon,~J.~et al.,~2018,~\mnras,~473,~4
\bibitem[{Kraft~et~al.~(2022)}]{lem} Kraft,~R.~et al.,~2020,~arXiv,~2211.09827
\bibitem[{Matthews~et~al.~(2022)}]{matthews22_bfields} Matthews,~J.~et al.,~2022,~\apj,~930,~1

\bibitem[{Halverson~et~al.~(2019)}]{halverson19} Halverson,~J.~et~al.,~2019,~PRL.~D,~100,~106010

\bibitem[{Schallmoser~et~al.~(2021)}]{schallmoser21} Schallmoser,~S.~et al.,~2021,~\mnras~541,~1

\bibitem[{Conlon~et~al.~(2017)}]{conlon17_manysources} Conlon,~J.~et al.,~2017,~JCAP~7,~5
\end{thebibliography}
\end{document}